\documentclass[journal]{IEEEtran}
\usepackage{algorithm,caption}
\usepackage[bottom]{footmisc}
\usepackage{multicol}
\usepackage{graphicx}
\usepackage{float}
\usepackage{amsmath}
\usepackage{amsfonts}
\usepackage{amssymb}
\usepackage{amsthm}
\usepackage{mathrsfs}
\usepackage{commath}
\usepackage[noadjust]{cite}
\usepackage{url}
\usepackage[all]{xy}
\usepackage{epsfig}
\usepackage{epstopdf}
\usepackage{amstext}
\usepackage{graphicx,tabularx}
\usepackage{graphics}
\usepackage{float}
\usepackage{multicol}
\usepackage{makeidx}
\usepackage{accents}
\usepackage[mathscr]{euscript}
\IEEEoverridecommandlockouts

\usepackage[noend]{algpseudocode}
\def\BState{\State\hskip-\ALG@thistlm}
\makeatother

\begin{document}
\title{\huge  Mode Selection Schemes for D2D Enabled Aerial Networks}
\author{
Aymen~Omri,~\IEEEmembership{Member,~IEEE,}
~Mazen~O.~Hasna,~\IEEEmembership{Senior~Member,~IEEE}
and~Muhammad~Zeeshan~Shakir,~\IEEEmembership{Senior~Member,~IEEE} 
\thanks{\scriptsize  A. Omri and M. O. Hasna are with the Department of Electrical Engineering, Qatar University, Qatar. E-Mail Address: omriaymen@qu.edu.qa, hasna@qu.edu.qa.}
\thanks{\scriptsize  M. Z. Shakir is with the School of Engineering and Computing, University
of the West of Scotland, Paisley, Scotland, UK, E-Mail Address: Muhammad.Shakir@uws.ac.uk}
\vspace{-0.4in}}

\maketitle
\begin{abstract}
In this paper, we present and evaluate the effect of two mode selection schemes for device to device (D2D) enabled areal netwroks. The two schemes are based on a threshold received signal strength (RSS) and an average threshold D2D distance between two given users to select the D2D mode. While one scheme triggers the D2D mode based on distance values only, the other scheme can trigger D2D mode for larger distances if a minimum RSS value is received, for it to maximize connectivity regions. Numerical results show the advantage of the presented schemes in offloading traffic from aerial platforms and the effect of the environment on the performance of D2D enabled aerial networks.
 \end{abstract}
\vspace{-0.05in}
\vspace{-0.15in}
\section{Introduction}
\vspace{-0.03in}
Network Flying Platforms (NFPs) such as drone and unmanned aerial vehicle (UAV) have been proposed as promising solutions for future cellular networks \cite{D2DNFP,D2DNFP1,D2DNFP2}. In particular, NFPs can be used to enhance the wireless capacity and expand the coverage for temporary events, where there is a high density of users in a limited area, e.g., sports events and concert gatherings. Furthermore, NFPs can be deployed for unexpected scenarios, such as in emergency situations to support disaster relief activities and to enable communications when conventional terrestrial networks are either damaged or crowded. In addition, owing to their mobility, NFPs can be quickly and efficiently deployed to support cellular networks and enhance network quality-of-service (QoS) \cite{D2DNFP,D2DNFP1}.

For the mentioned scenarios, where there is no infrastructure or it is difficult to deploy one due to limited resources/support, direct device-to-device (D$2$D) communication helps in offloading traffic from NFP’s which is critical in many scenarios, especially when NFPs are assisting the ground users in maintaining their links and in establishing the D2D links. Considerable recent work have studied the use of D$2$D communications between wireless users over the licensed spectrum \cite{MS3}. Different from NFP-based communications, the D$2$D transmission allows sources and target destinations which are close to each other to directly communicate without being relayed by the NFPs in the network. However, when the users are far from each other, the NFP-based aerial network may outperforms the D$2$D communication. As a result, proposing efficient mode selection schemes for D$2$D enabled aerial networks is of high importance.

To the best of our knowledge, only few papers have addressed the analysis of mode selection in D$2$D enabled aerial networks. In \cite{D2DNFP}, the authors have analyzed the coverage and rate performance of UAV-based wireless communication in the presence of underlaid D2D communication links. The advantage of alternative connectivity options, such as D2D links and drone-assisted access, have been evaluated in \cite{D2DNFP1}, where the authors have confirmed improvements of up to 40 percent in link availability and reliability with the use of proximate connections on top of the cellular-only baseline. In \cite{D2DNFP2}, a drone-assisted multi-hop device-to-device (D2D) communication scheme has been proposed as a means to extend the network coverage over regions where it is difficult to deploy a terrestrial relay. All the above work have been focusing on evaluating the performance of D$2$D enabled aerial network without proposing mode selection techniques, which is a critical issue for D2D-enabled mobile networks in general.

In light of the aforementioned related work, our main contributions can be summarized as follows:

\begin{itemize}
\item  We propose two mode selection schemes for D2D enabled aerial networks. To reduce the signaling overhead between the different nodes in the network, the proposed schemes use predefined threshold received signal strength (RSS) and an average threshold D2D distance between a given two users, where the D2D mode can be established. More details about the proposed mode selections are presented in Section III.\\ \vspace{-0.15in}

\item To describe the system model, and different than the work in \cite{D2DNFP,D2DNFP1,D2DNFP2}, stochastic geometry is used in this paper, where we derive the expressions of the average threshold D$2$D distance, and the probability of using D$2$D mode to evaluate and investigate the two mode selection schemes. \\ \vspace{-0.1in}
\end{itemize}

\vspace{-0.2in}
\section{System Model}
\vspace{-0.1in}

	\begin{figure}[b!]		
	\vspace{-0.3in}
    \centering
        \includegraphics[width=8.5cm]{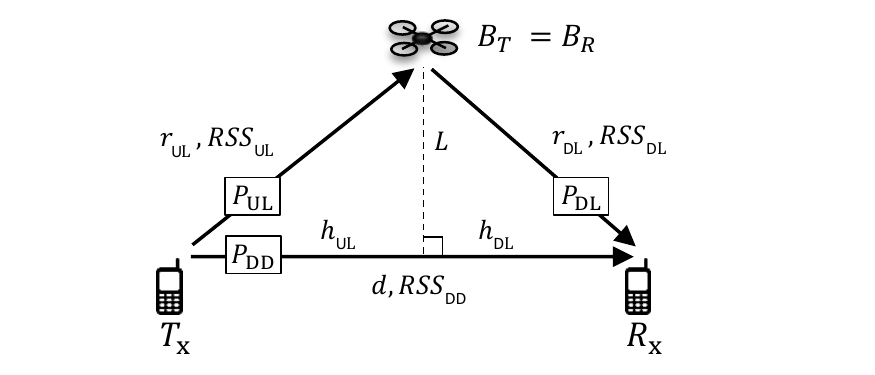}
				    \vspace{-0.15in}
    \caption{The system model.}
\end{figure}

\subsection{Network Model}
We consider a flying platform based wireless communication network. The network consists of randomly located users, and a number of flying platforms as shown in Fig. 1. The users are randomly located in the network, where the active transmitters are distributed according to a Poisson point process (PPP) $\Phi_{_{{T}}}$ with density $\lambda_{_{{T}}}$, and the receivers are located according to a PPP $\Phi_{_{{R}}}$ with density $\lambda_{_{{R}}}$. The locations of the flying platforms are assumed to be distributed according to a Mat\'ern hard-core process (MHCP) of Type II $\Phi_{_{{B}}}$, with density $\lambda_{_{{B}}}$, and a minimum inter-distance $\delta$ between each two platforms\footnote{We are not considering the NFP placement problem in this paper, but rather we are looking at a snapshot of the network following the placement of NFPs.}. $\Phi_{_{{B}}}$ is based on a parent PPP $\Phi_{_{{P}}}$ of density $\lambda_{_{{P}}}$. In MHCP of type II process, each parent nodes has a random associated mark, and a node is not retained only if there is another node within distance $\delta$ with a smaller random associated mark \cite{Haenggi-MHCP}. Accordingly, the new density is expressed as \cite{Haenggi-MHCP}

\vspace{-0.1in} {\footnotesize 
\begin{equation} \label{SM2}
\lambda_{_{{B}}} = \tau \lambda_{_{{P}}} = \frac{1- \exp (-\pi \lambda_{_{{P}}} \delta^{^2} )}{\pi \delta^{^2}}.
\end{equation} } \vspace{-0.4in}

\subsection{Transmission Modes}
Based on the presented network model, a given transmitter $T_{_{\text{x}}}$ and a given receiver $R_{_{\text{x}}}$ can communicate using one of the following considered modes:

\subsubsection{Standard Transmission Mode}
Within this mode, $T_{_{\text{x}}}$ and $R_{_{\text{x}}}$ communicate with the corresponding nearest platforms for the down-link (DL) and up-link (UL), where we assume that the UL and DL are decoupled, and that the platforms have error free channel between them. Let ${B}_{_{{T}}}$ and ${B}_{_{{R}}}$ denote the nearest platforms to $T_{_{\text{x}}}$ and $R_{_{\text{x}}}$, respectively, with the corresponding distances are denoted by $r_{_{\text{UL}}}$ and $r_{_{\text{DL}}}$. Let  $h\ \in \{h_{_{\text{UL}}}, h_{_{\text{DL}}}\}$ be the horizontal distance between a given user and its nearest platform, and $L$ be the altitude of the platforms as presented in Fig. 1. The path-loss model for air-to-ground communications can be presented as follows \cite{Ref-PLoss-1}

\vspace{-0.17in} {\footnotesize 
\begin{align} \label{SM3}
PL_{_{\text{AtG}}}(h,L)  & =  20\ \log_{_{{10}}}{\left(\frac{4\pi f_{_{{c}}}}{C}\right)} + 20\ \log_{_{{10}}}{(r)} \nonumber \\
& + {P}_{_{\text{LOS}}}(h,L)\ \eta_{_{\text{LOS}}} + [1-{P}_{_{\text{LOS}}}(h,L)]\ \eta_{_{\text{NLOS}}},
\end{align}} \vspace{-0.2in}

\hspace{-0.15in}where, $f_{_{{c}}}$ is the carrier frequency [Hz], $C$ is the speed of light [m/s], $\eta_{_{\text{LOS}}}$ and $\eta_{_{\text{NLOS}}}$ [dB] are the losses (depending on the environment of the Line of Sight (LOS) and non-LOS connections, respectively, and ${P}_{_{\text{LOS}}}(h,L)$ is the LOS probability, which is expressed as follows

\vspace{-0.2in} {\footnotesize 
\begin{align} \label{SM4}
{P}_{_{\text{LOS}}}(h,L)  =  \frac{1}{1+a \exp{\Big(-b \Big[\arctan{\Big(\frac{L}{h}\Big)}-a\Big]\Big)}},
\end{align} } \vspace{-0.1in}

\hspace{-0.15in}where, $a$ and $b$ are constant parameters that depend on the environment.
Based on that, the value of the air-to-ground signal attenuation can be presented as $ A(h,L)\ r^{^{-2}}$, where $A(h,L)$ is expressed as

\vspace{-0.12in} {\footnotesize 
\begin{equation} \label{SM5}
A (h,L) = \Big(\frac{C}{4\pi f_{_{{c}}}}\Big)^{^2} \ 10^{^{-\frac{{P}_{_{\text{LOS}}}(h,L)(\eta_{_{\text{LOS}}}-\eta_{_{\text{NLOS}}})+\eta_{_{\text{NLOS}}}}{10}}},
\end{equation} } \vspace{-0.15in}

\hspace{-0.15in}and, $r = \sqrt{h^{^2}+L^{^2}}$ is the distance between a given user and its serving nearest platform.\\
\vspace{-0.1in}
\subsubsection{D$2$D Transmission Mode}
This mode enables the users to communicate directly without going through the platform, where the NFP role here is to facilitate and help establishing  the D2D connection by exchanging the control and coordination signals. Let $d$ be the distance, in meters, between $T_{_{\text{x}}}$ and $R_{_{\text{x}}}$. The path-loss model for D2D communications is given by

\vspace{-0.17in} {\footnotesize 
\begin{align} \label{SM55}
PL_{_{\text{DD}}}(d)  & =  20\ \log_{_{{10}}}{\left(\frac{4\pi f_{_{{c}}}}{C}\right)} + 10\ \alpha \log_{_{{10}}}{(d)},
\end{align}} \vspace{-0.2in}

\hspace{-0.15in}where $\alpha$ is the path Loss exponent. Consequently, the D$2$D signal attenuation can be presented as $ A_{_{\text{DD}}}\ d^{^{-\alpha}}$, with $
A_{_{\text{DD}}} = (C/ [4\pi f_{_{{c}}}] )^{^2}$.

\vspace{-0.15in}
\section{Proposed Schemes Description}
In this section, we propose two new communication schemes for D2D-enabled aerial networks. For the different scenarios, the mode selection method is considering the RSSs of the D$2$D and aerial links. The measurement results of RSS
are usually averaged with filtering the received signal out through Layer-1 and Layer-3 \cite{Ref-mselcet}. Therefore, only the distance based path-loss is
considered in RSS modeling \cite{Ref-mselcet}. According to the power law, the RSSs for the D2D link, the DL, and the UL are presented as follows

\vspace{-0.1in} {\footnotesize 
\begin{equation} \label{PS1}
RSS_{_{\text{DD}}} = P_{_{\text{DD}}}\ A_{_{\text{DD}}} \ d^{^{-\alpha}}, 
\end{equation}}\vspace{-0.1in}
{\footnotesize 
\begin{equation} \label{PS3}
RSS_{_{\text{DL}}} = P_{_{\text{DL}}}\ A_{_{\text{DL}}} \ r_{_{\text{DL}}}^{-2}, 
\end{equation} } \vspace{-0.25in}

\hspace{-0.15in}and

\vspace{-0.2in}{\footnotesize \begin{equation} \label{PS2}
RSS_{_{\text{UL}}} = P_{_{\text{UL}}}\ A_{_{\text{UL}}} \ r_{_{\text{UL}}}^{-2}, 
\end{equation} } \vspace{-0.2in}

\hspace{-0.15in}respectively, where $P_{_{\text{DD}}}$ is the transmit power of $T_{_{\text{x}}}$ when the D$2$D mode is used, $P_{_{\text{UL}}}$ and $P_{_{\text{DL}}}$ are the transmit powers of $T_{_{\text{x}}}$ and $R_{_{\text{x}}}$, respectively, when the standard mode is used, $A_{_{\text{UL}}} = A(h_{_{\text{UL}}},L)$, and $A_{_{\text{DL}}} = A(h_{_{\text{DL}}},L)$. Accordingly, the selection mode in the proposed schemes depends mainly on the threshold D$2$D distance, which can be defined as the maximum distance between $T_{_{\text{x}}}$ and $R_{_{\text{x}}}$, where the D$2$D mode can be used. Let $d_{_{\text{th}}}(L)$ be this distance that satisfies $RSS_{_{\text{DD}}} =  \text{min}(RSS_{_{\text{UL}}}, RSS_{_{\text{DL}}})$.  Based on (\ref{PS1}), $d_{_{\text{th}}}(L)$ can be expressed as 

\vspace{-0.27in} {\footnotesize 
\begin{align} \label{PS4}
d_{_{\text{th}}}(L)  = \Bigg[\frac{P_{_{\text{DD}}}\ A_{_{\text{DD}}}}{\text{min}(RSS_{_{\text{UL}}}, RSS_{_{\text{DL}}})}\Bigg]^{^{\frac{1}{\alpha}}}.
\end{align} } \vspace{-0.15in}

\hspace{-0.15in}Furthermore, and based on (\ref{PS1}-\ref{PS4}), the expression of $d_{_{\text{th}}}(L)$ can be rewritten as follows

\vspace{-0.15in} {\footnotesize 
\begin{align} \label{PS5}
d_{_{\text{th}}}(L)  = \begin{cases}
\Big[\frac{P_{_{\text{DD}}} \ A_{_{\text{DD}}} }{ P_{_{\text{UL}}} \ A_{_{\text{UL}}}}\Big]^{^{\frac{1}{\alpha}}}\ r_{_{\text{UL}}}^{^{\frac{2}{\alpha}}}, &\ \ \mbox{if } \  r_{_{\text{DL}}} \leq \sqrt{\frac{P_{_{\text{DL}}} \ A_{_{\text{DL}}} }{P_{_{\text{UL}}} \ A_{_{\text{UL}}}}}\ r_{_{\text{UL}}}\\
\Big[\frac{P_{_{\text{DD}}} \ A_{_{\text{DD}}} }{ P_{_{\text{DL}}} \ A_{_{\text{DL}}}}\Big]^{^{\frac{1}{\alpha}}}\ r_{_{\text{DL}}}^{^{\frac{2}{\alpha}}}, &\ \ \mbox{if } \  r_{_{\text{UL}}} \leq \sqrt{\frac{P_{_{\text{UL}}} \ A_{_{\text{UL}}} }{P_{_{\text{DL}}} \ A_{_{\text{DL}}}}}\ r_{_{\text{DL}}}. 
\end{cases}
\end{align} } \vspace{-0.15in}

\hspace{-0.15in}To reduce the signaling overhead between the different nodes, which is used to update the different instantaneous RSS values, the proposed mode selection schemes are based on the average of $d_{_{\text{th}}}(L), \bar{d}_{_{\text{th}}}(L)$, instead of its instantaneous value. \\
To derive $\bar{d}_{_{\text{th}}}(L)$, the probability density function (PDF) expression of the distance $r$ between a user and a serving nearest platform, denoted by $f$, is needed. The exact expression of $f$ is derived in \cite{Ref-PDF} and is expressed as follows\\

\vspace{-0.3in} {\scriptsize
\begin{align} \label{PDF1}
 & f(r)  = 2 \lambda_{_{{P}}} \pi  \sqrt{r^{^2}-L^{^2}} \ \left[\hspace{-0.02in} \frac{1-\exp{\Big(-\lambda_{_{{P}}}\Big[\pi \delta^{^2}-g\Big(\sqrt{r^{^2}-L^{^2}},\delta\Big)\Big]\Big)}}{\lambda_{_{{P}}}\Big[\pi \delta^{^2}-g\Big(\sqrt{r^{^2}-L^{^2}},\delta\Big)\Big]}\right]\nonumber \\
&\times \exp \hspace{-0.02in}\left(\hspace{-0.02in} - \hspace{-0.02in}\int_{L}^{r} \hspace{-0.02in} 2 \pi y \lambda_{_{{P}}} \left[\hspace{-0.02in} \frac{1-\exp{\Big(-\lambda_{_{{P}}}\Big[\pi \delta^{^2}-g\Big(\sqrt{y^{^2}-L^{^2}},\delta\Big)\Big]\Big)}}{\lambda_{_{{P}}}\Big[\pi \delta^{^2}-g\Big(\sqrt{y^{^2}-L^{^2}},\delta\Big)\Big]}\right] \hspace{-0.02in} dy \hspace{-0.02in}\right) \hspace{-0.02in} .
\end{align} } \vspace{-0.15in}

\hspace{-0.15in}where,

\vspace{-0.25in} {\footnotesize 
\begin{align} \label{CDF3}
g(x,\delta)  = \begin{cases}
2\ \delta^{^2} \cos^{^{-1}}\big(\frac{x}{2\ \delta}\big)-\ \frac{1}{2}\ x\ \sqrt{4\ \delta^{^2} - x^{^2}}, \ \ \ \ \ \ \mbox{if} \  0< x \leq 2\ \delta& \\
& \\
 0 , \hspace{1.85in} \ \ \ \ \ \mbox{if} \  x > 2\ \delta. &
\end{cases}
\end{align}} \vspace{-0.15in}

\hspace{-0.15in}Due to the complicated exact expressions in (\ref{PDF1}), a tight approximate PDF expression, which is based on the work in \cite{Aymen-CL1}, is used to derive tractable analytical results for the proposed mode selection schemes. This approximate expression is given by

\vspace{-0.25in} {\footnotesize 
\begin{align}\label{PDF2}
f(r)  \approx 2 \lambda_{_{{B}}} \pi r \ \exp{\Big(- \lambda_{_{{B}}} \pi  \Big[r^{2} - L^{^2}\Big]\Big)}.
\end{align}} \vspace{-0.2in}

\hspace{-0.14in}To confirm the accuracy of the approximate expression in (\ref{PDF2}), we present in Fig. 2 the histogram for simulating the PDF of $r$ as well as the outputs of the exact and approximate PDF expressions in (\ref{PDF1}) and (\ref{PDF2}), respectively. As shown in this figure, there is an excellent fit between the exact and the approximate expressions, where an average error of less than $2$e$-5$ ($1.678e-05$) is observed. Now, by using the PDF expression in (\ref{PDF2}), and the expression of $d_{_{\text{th}}}(L)$ in (\ref{PS5}), the final expression of $\bar{d}_{_{\text{th}}}(L)$ is derived in Appendix A and is given by

\vspace{-0.2in} {\footnotesize 
\begin{align} \label{dth}
&\bar{d}_{_{\text{th}}}(L) \approx 2 \Bigg[\frac{P_{_{\text{DD}}} A_{_{\text{DD}}} }{\pi  \lambda_{_{{B}}}P_{_{\text{UL}}} \tilde{A}}\Bigg]^{^{\frac{1}{\alpha}}} \Bigg[\hspace{-0.02in}\exp{\Big(\hspace{-0.02in} \lambda_{_{{B}}} \pi L^{^2}\Big)}\Gamma\bigg(\frac{\alpha+1}{\alpha},\lambda_{_{{B}}} \pi L^{^2}\bigg) \nonumber \\ 
&\hspace{-0.02in}-\exp{\Big(2\lambda_{_{{B}}} \pi L^{^2}\Big)} \hspace{-0.02in} \Bigg[\frac{P_{_{\text{UL}}}}{P_{_{\text{UL}}}\hspace{-0.02in}+\hspace{-0.02in}P_{_{\text{DL}}}}\Bigg]^{^{\frac{\alpha+1}{\alpha}}}\Gamma\Bigg(\frac{\alpha+1}{\alpha},\Bigg[\frac{P_{_{\text{UL}}}\hspace{-0.02in}+\hspace{-0.02in}P_{_{\text{DL}}}}{P_{_{\text{UL}}}}\Bigg] \lambda_{_{{B}}} \pi L^{^2}\Bigg)\hspace{-0.02in} \Bigg].
\end{align} }\vspace{-0.2in}

\hspace{-0.15in}The approximation in (\ref{dth}) is due to using the approximate expression $\tilde{A}\ (= A(1/[2\sqrt{\lambda_{_B}}],L)$ of $A_{_{\text{UL}}}$ and $A_{_{\text{DL}}}$, and the approximate expression of $f$ in (\ref{PDF2}).

\begin{figure}[t!]
  \centering
        \includegraphics[width=7cm]{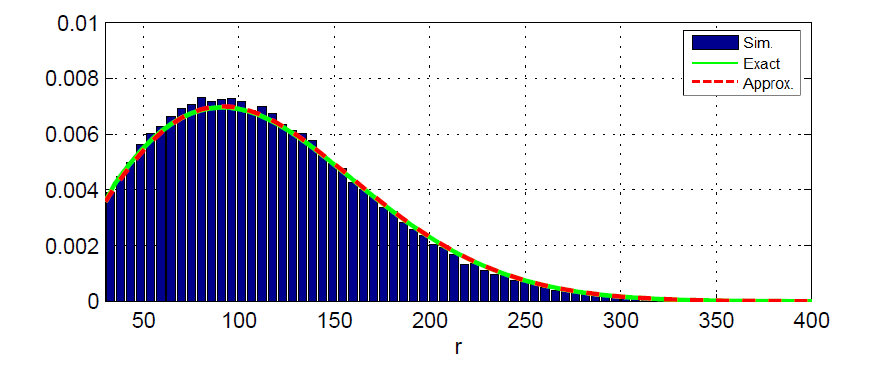}
				\vspace{-0.15in}
				   \caption{\footnotesize Histogram for the PDF simulation of $r$ as compared to the exact and approximate expressions in (\ref{PDF1}) and (\ref{PDF2}), respectively, with $\lambda_{_{{P}}}=2e-5/$m$^{2}$, $R=500$ m, and $\delta=100$ m.}
					\vspace{-0.25in}
\end{figure}

Consequently, based on the average threshold D$2$D distance, we propose and evaluate the performance of the following transmission schemes.
\vspace{-0.15in}
\subsection{Threshold D2D Distance based Scheme (TDDS)}
In this scheme, $T_{_{\text{x}}}$ can transmit its message ${M}$ to $R_{_{\text{x}}}$ by using either the D$2$D or standard transmission modes. The TDDS algorithm can be summarized as follows

\vspace{-0.1in}
\begin{algorithm}
\caption*{\footnotesize TDDS Algorithm }\label{euclid}
\begin{algorithmic}[1]
\State {\footnotesize ${M} \neq \emptyset$, $d = r_{_0}$ : The initial distance between $T_{_{\text{x}}}$ and $R_{_{\text{x}}}$}
\While {\footnotesize (${M} \neq \emptyset$))}

    \If {\footnotesize ($d \leq \bar{d}_{_{\text{th}}}(L)$)} 
        \State {\footnotesize The D2D mode will be used.}
    \Else
	      	\If {\footnotesize ($T_{_{\text{x}}}$ is associated)}
	      	  	\State {\footnotesize The standard transmission mode will be used.}
		      \Else 
					    \If {\footnotesize ($RSS_{_{\text{DD}}} \geq RSS_{_{\text{th}}}$)}
	      	  	\State {\footnotesize The D2D mode will be used.}
					    \EndIf
							\State {\textbf{Endif}}
					\EndIf
					\State {\textbf{Endif}}
		\EndIf
	\State {\textbf{Endif}}
	\State {\footnotesize Updating $d$.}
\State \hspace{-0.3in} {\textbf{Endwhile}}
\EndWhile
\end{algorithmic}
\end{algorithm}

\vspace{-0.1in} 

\hspace{-0.15in}Within this scheme, the mode selection is mainly based on the predefined values of $\bar{d}_{_{\text{th}}}(L)$ and $RSS_{_{\text{th}}}$, and the actual distance $d$ between $T_{_{\text{x}}}$ and $R_{_{\text{x}}}$. The distance $d$ is assumed to be available at the corresponding nodes, which can update its value periodically. Based on that, if $d \leq \bar{d}_{_{\text{th}}}(L)$, then the  D2D mode is used, where the users are assisted by the NFPs for D$2$D link establishment. Otherwise, and given that $T_{_{\text{x}}}$ is associated with a platform, the standard transmission mode is used. If $T_{_{\text{x}}}$ is not associated, then the D2D mode is used if $RSS_{_{\text{DD}}} \geq RSS_{_{\text{th}}}$, which gives the possibility to the non associated users to directly communicate using a predefined $RSS_{_{\text{th}}}$. Here, the signaling overhead between the different nodes can be reduced, as the proposed scheme is based only on the predefined parameters $\bar{d}_{_{\text{th}}}(L)$, $RSS_{_{\text{th}}}$, and the actual distance $d$. In summary, the D2D mode for this scheme is used in two events: the first is when the distance $d$ between $T_{_{\text{x}}}$ and $R_{_{\text{x}}}$ is less than or equal to $\bar{d}_{_{\text{th}}}(L)$. The second event is when $d$ is larger than $\bar{d}_{_{\text{th}}}(L)$, $T_{_{\text{x}}}$ is not associated, and $RSS_{_{\text{DD}}}$ is larger or equal to $RSS_{_{\text{th}}}$. Accordingly, the probability of using D2D mode can be written as follows

\vspace{-0.15in} {\footnotesize 
\begin{align}
&P_{_{\text{D2D}}}^{^{\text{TDDS}}}  = Pr \Big\{d \leq \bar{d}_{_{\text{th}}}(L)\Big\} \nonumber \\
        & + Pr\Big\{ \big(d > \bar{d}_{_{\text{th}}}(L)\big) \cap \big(T_{_{\text{x}}} \text{is not associated}\big) \cap \big(RSS_{_{\text{DD}}} \geq RSS_{_{\text{th}}}\big)\Big\}.
\end{align} }\vspace{-0.15in}

\hspace{-0.15in}As presented in the system model, the users (including $T_{_{\text{x}}}$ and $R_{_{\text{x}}}$) are randomly distributed according to PPPs. By considering the transmitter $T_{_{\text{x}}}$, its target receiver $R_{_{\text{x}}}$ can be located at any point in the total considered area ($\pi R^{^2}$). Based on that, the expression of $P_{_{\text{D2D}}}^{^{\text{TDDS}}} $ can be derived as 

\vspace{-0.2in}
 {\footnotesize
\begin{align} 
P_{_{\text{D2D}}}^{^{\text{TDDS}}}& \hspace{-0.04in} = \hspace{-0.03in} Pr\Big\{d \leq \bar{d}_{_{\text{th}}}(L)\Big\} \hspace{-0.02in}+\hspace{-0.02in}\Big(1\hspace{-0.02in}-\hspace{-0.02in}p\Big) Pr\hspace{-0.02in}\left\{\hspace{-0.02in}\bar{d}_{_{\text{th}}}(L) \hspace{-0.03in}< \hspace{-0.02in} d \hspace{-0.02in}\leq \hspace{-0.04in} \left( \hspace{-0.02in} \bar{r}_{_{\text{th}}} \hspace{-0.04in} = \hspace{-0.04in} \Big[\frac{P_{_{\text{DD}}} A_{_{\text{DD}}}}{RSS_{_{\text{th}}}}\Big]^{^{\frac{1}{\alpha}}}\hspace{-0.04in} \right) \hspace{-0.06in} \right\}  \nonumber \\
& = \begin{cases}
\frac{\bar{d}_{_{\text{th}}}(L)^{^2}}{R^{^2}} + \frac{\big(\bar{r}_{_{\text{th}}}^{^2}-\bar{d}_{_{\text{th}}}(L)^{^2}\big)\ \big(1-p\big)\ }{R^{^2}}, \ \ \mbox{if} \ \ \ \ \ \ (\bar{d}_{_{\text{th}}}(L) \leq \bar{r}_{_{\text{th}}} \leq R)& \\
\frac{\bar{d}_{_{\text{th}}}(L)^{^2}}{R^{^2}} + \big(1-\frac{\bar{d}_{_{\text{th}}}(L)^{^2}}{R^{^2}}\big) \big(1-p\big)\ , \  \mbox{elseif} \ \ (\bar{d}_{_{\text{th}}}(L) \leq R < \bar{r}_{_{\text{th}}} ), & \\
\frac{\bar{d}_{_{\text{th}}}(L)^{^2}}{R^{^2}} , \ \ \ \ \ \ \ \ \ \ \ \ \ \ \ \ \ \ \ \ \ \ \ \ \ \ \ \ \ \ \ \ \ \mbox{elseif} \ \ (\bar{r}_{_{\text{th}}} \leq \bar{d}_{_{\text{th}}}(L) \leq R)& \\
1 , \ \ \ \ \ \ \ \ \ \ \ \ \ \ \ \ \ \ \ \ \ \ \ \ \ \ \ \ \ \ \ \ \ \ \ \ \ \ \ \ \ \mbox{else}.& \\
\end{cases}
\end{align} }\vspace{-0.35in}

\vspace{-0.1in}
\subsection{Received Signal Strength based Scheme (RSSS)}

Similar to the TDDS algorithm, the mode selection in this scheme is based on the predefined values of $\bar{d}_{_{\text{th}}}(L)$, $RSS_{_{\text{th}}}$, and the actual distance $d$. However, D2D mode is used only in the event in which $RSS_{_{\text{DD}}}$ is larger or equal to $RSS_{_{\text{th}}}$. In the case of a large actual value of $d$, when $RSS_{_{\text{DD}}} < RSS_{_{\text{th}}}$ and $d > \bar{d}_{_{\text{th}}}(L)$, the standard transmission mode may be used, as it outperforms the D2D mode in this case as shown in the following algorithm.

\vspace{-0.1in}
\begin{algorithm}
\caption*{\footnotesize RSSS Algorithm}\label{euclid}
\begin{algorithmic}[1]
\State {\footnotesize ${M} \neq \emptyset$, $d = r_{_0}$ : The initial distance between $T_{_{\text{x}}}$ and $R_{_{\text{x}}}$}
\While {\footnotesize (${M} \neq \emptyset$))}
    \If {\footnotesize ($RSS_{_{\text{DD}}} \geq RSS_{_{\text{th}}}$)}
	      	  	\State {\footnotesize The D2D mode will be used.}
	  \Else
        \If {\footnotesize ($d > \bar{d}_{_{\text{th}}}(L)$)} 
          	\If {\footnotesize ($T_{_{\text{x}}}$ is associated)}
	      	      	\State {\footnotesize The standard transmission mode will be used.}
		        \EndIf
			    	\State {\textbf{Endif}}
		    \EndIf
		    \State {\textbf{Endif}}
		\EndIf
		\State {\textbf{Endif}}
  	\State {\footnotesize Updating $d$.}
\State \hspace{-0.3in} {\textbf{Endwhile}}
\EndWhile
\end{algorithmic}
\end{algorithm}
\vspace{-0.13in}

\hspace{-0.15in}This scheme can be used for emergency communications, e.g., during disaster, when a minimum small value of $RSS_{_{\text{th}}}$ is required to use the D2D mode. Consequently, the probability of using D2D mode, within this scheme, is expressed as  

\vspace{-0.15in} {\footnotesize 
\begin{align} 
P_{_{{D2D}}}^{^{\text{RSSS}}} & = Pr\Big\{RSS_{_{\text{DD}}} \geq RSS_{_{\text{th}}}\Big\} = Pr\left\{d \leq \bar{r}_{_{\text{th}}}\right\}\nonumber \\
&  = \begin{cases}
\frac{\bar{r}_{_{\text{th}}}^{^2}}{R^{^2}}, \ \ \ \mbox{if} \ \  \Big(RSS_{_{\text{th}}} \geq \frac{P_{_{\text{DD}}}\ A_{_{\text{DD}}}} {R^{^2}} \Big)& \\
1,\ \ \ \ \ \mbox{else}. &
\end{cases}
\end{align} }

\vspace{-0.27in}
\section{Numerical results and Discussions}
In this section, numerical results are presented to investigate the performance of the proposed schemes in terms of the probability of using D2D mode. Fig. 3 presents the probability of using D2D mode vs. $L$ for TDDS and RSSS algorithms. As shown in this figure, for RSSS, $P_{_{{D2D}}}$ is invariable with the increased values of $L$, which is not the case for TDDS. This is due to the fact that the probability of using D2D mode within RSSS depends only on the values of $RSS_{_{\text{th}}}$. However, for TDDS, this probability is not only inversely proportional to $RSS_{_{\text{th}}}$ but also depends on the height of the NFP $L$. In fact, and based on Fig. 3, at low altitudes ($<300$ m), $P_{_{{D2D}}}$ decreases as $L$ increases due to the increased LOS for both the UL and DL. However, if the height increases further, we notice that the D2D probability increases as the path loss component for the standard communication mode starts to overcome the LOS component. This becomes more clear as the height increases further, beyond $2$ km for the shown parameters where $P_{_{{D2D}}}$ becomes dominant and approaches 1. In Fig. 4, the variation of $P_{_{{D2D}}}$ vs. $L$ is presented for different types of environments. As shown in the figure, and for $ L<2$ km, $P_{_{{D2D}}}$ is higher for less dense environments. This is expected, as the LOS probability for D2D mode increases significantly with the decreased obstacles, when compared to the LOS probability for the standard mode. Consequently, an important enhancement of the D2D RSS can be observed, which results in an increase of the $P_{_{{D2D}}}$. Again, at relatively higher altitudes, D2D communication is dominant in all types of environments due to the severe path loss component of the standard communication.\\
The presented results confirm the advantage of using the proposed schemes in offloading traffic from NFPs. Indeed, for a given environment parameters, a range of the NFP height can be selected to maximize the probability of using D2D mode, which gives better link reliability than the standard communication in this range.

\begin{figure}[t!]
  \centering
        \includegraphics[width=7cm,height=5cm]{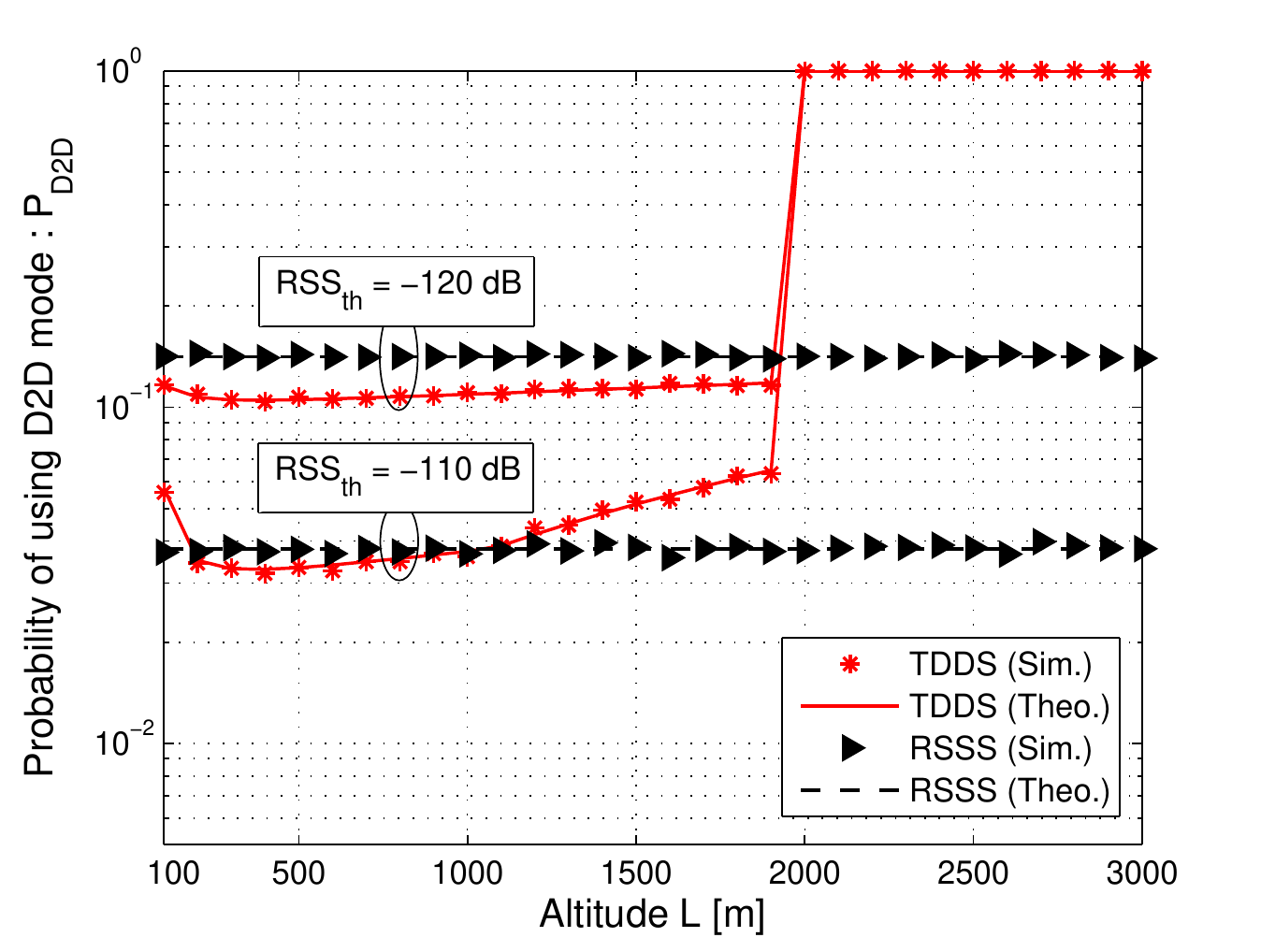}
				\vspace{-0.1in}
				    \caption{\footnotesize $P_{_{{D2D}}}$ vs. $L$, in a highrise urban environment, with $f_{_{c}} = 2.5$ GHz, $R = 500$ m, $\lambda_{_{{B}}} = 1e-4$ m$^{-2}$, $\delta = 100$ \ m, $\lambda_{_{{T}}} = 1e-3$ m$^{-2}$, $\alpha = 3.5$, $a=27.23$, $b=0.08$, $\eta_{_{\text{LOS}}} = 2.3$ dB, $\eta_{_{\text{NLOS}}} = 34$ dB, and different values of $RSS_{_{\text{th}}}$ .}
						\vspace{-0.2in}
\end{figure}

	\vspace{-0.15in}
\appendices
\section{}
\vspace{-0.05in}

Based on (\ref{PS5}), the expression of $\bar{d}_{_{\text{th}}}(L)$ can be written as

\vspace{-0.1in}{\scriptsize
\begin{align} \label{A1}
 \bar{d}_{_{\text{th}}}(L) & =  \int_{L}^{\infty}\int_{L}^{\sqrt{\frac{P_{_{\text{DL}}} \ A_{_{\text{DL}}} }{P_{_{\text{UL}}} \ A_{_{\text{UL}}}}} r_{_{\text{UL}}}} \ \Bigg[\frac{P_{_{\text{DD}}} \ A_{_{\text{DD}}} }{ P_{_{\text{UL}}} \ A_{_{\text{UL}}}}\Bigg]^{^{\frac{1}{\alpha}}} r_{_{\text{UL}}}^{^{\frac{2}{\alpha}}} \ f(r_{_{\text{UL}}})\ f(r_{_{\text{DL}}}) dr_{_{\text{DL}}} dr_{_{\text{UL}}} \nonumber \\
& + \int_{L}^{\infty}\int_{L}^{\sqrt{\frac{P_{_{\text{UL}}} \ A_{_{\text{UL}}} }{P_{_{\text{DL}}} \ A_{_{\text{DL}}}}} r_{_{\text{DL}}}} \ \ \Bigg[\frac{P_{_{\text{DD}}} \ A_{_{\text{DD}}} }{ P_{_{\text{DL}}} \ A_{_{\text{DL}}}}\Bigg]^{^{\frac{1}{\alpha}}} r_{_{\text{DL}}}^{^{\frac{2}{\alpha}}} \ f(r_{_{\text{UL}}})\ f(r_{_{\text{DL}}}) dr_{_{\text{UL}}} dr_{_{\text{DL}}}.
\end{align}}\vspace{-0.2in}

\hspace{-0.15in}A tight approximation of the expressions of $A_{_{\text{UL}}}$ and $A_{_{\text{DL}}}$, denoted by $\tilde{A}$, can be used by substituting $h$ with its mean value, $\bar{h}=\frac{1}{2\sqrt{\lambda_{_{{B}}}}}$, in (\ref{SM5}). Based on that, and by using the expressions of $f$ in (\ref{PDF2}), the integration with respect to $r_{_{\text{DL}}}$, in the first part, $I_1$, of the integrations in (\ref{A1}), is evaluated as follows

\vspace{-0.15in}{\scriptsize
\begin{align} \label{A2}
 I_1 & = \int_{L}^{\infty}\int_{L}^{\sqrt{\frac{P_{_{\text{DL}}} \ A_{_{\text{DL}}} }{P_{_{\text{UL}}} \ A_{_{\text{UL}}}}} r_{_{\text{UL}}}} \ \Bigg[\frac{P_{_{\text{DD}}} \ A_{_{\text{DD}}} }{ P_{_{\text{UL}}} \ A_{_{\text{UL}}}}\Bigg]^{^{\frac{1}{\alpha}}}\ r_{_{\text{UL}}}^{^{\frac{2}{\alpha}}} \ f(r_{_{\text{UL}}})\ f(r_{_{\text{DL}}})\ dr_{_{\text{DL}}}\ dr_{_{\text{UL}}} \nonumber \\
 & \approx  \Bigg[\frac{P_{_{\text{DD}}} \ A_{_{\text{DD}}} }{ P_{_{\text{UL}}} \ \tilde{A}}\Bigg]^{^{\frac{1}{\alpha}}} \int_{L}^{\infty} \Bigg[1-\exp{\Bigg(- \lambda_{_{{B}}} \pi \Bigg[\frac{P_{_{\text{DL}}}}{P_{_{\text{UL}}}}\ r_{_{\text{UL}}}^{2}-L^{^2}\Bigg]\Bigg)} \Bigg] \nonumber \\
& \times \Bigg[2 \lambda_{_{{B}}} \pi\ r_{_{\text{UL}}}^{^{1+\frac{2}{\alpha}}} \ \exp{\Bigg(- \lambda_{_{{B}}} \pi \Bigg[r_{_{\text{UL}}}^{2}-L^{^2}\Bigg]\Bigg)} \Bigg] d r_{_{\text{UL}}}
\end{align}}\vspace{-0.15in}

\hspace{-0.15in}By using  the change in variables: $x = \lambda_{_{{B}}} \pi r_{_{\text{UL}}}^{2}$, and integration by parts \cite[Eq. (3.381.3)]{Ref14}, the integration in (\ref{A2}) is evaluated, which yields to 

\vspace{-0.2in}{\scriptsize
\begin{align} \label{A3}
 I_1 & \approx   \Bigg[\frac{P_{_{\text{DD}}} A_{_{\text{DD}}} }{\pi  \lambda_{_{{B}}}P_{_{\text{UL}}} \tilde{A}}\Bigg]^{^{\frac{1}{\alpha}}}\hspace{-0.04in}\exp{\Big(2\lambda_{_{{B}}} \pi L^{^2}\Big)} \hspace{-0.02in} \Bigg[\hspace{-0.02in}\exp{\Big(\hspace{-0.02in}-\hspace{-0.02in} \lambda_{_{{B}}} \pi L^{^2}\Big)} \Gamma\bigg(\frac{\alpha+1}{\alpha},\lambda_{_{{B}}} \pi L^{^2}\bigg) \nonumber \\
& -\Bigg[\frac{P_{_{\text{UL}}}}{P_{_{\text{UL}}}+P_{_{\text{DL}}}}\Bigg]^{^{\frac{\alpha+1}{\alpha}}}\Gamma\Bigg(\frac{\alpha+1}{\alpha},\Bigg[\frac{P_{_{\text{UL}}}+P_{_{\text{DL}}}}{P_{_{\text{UL}}}}\Bigg] \lambda_{_{{B}}} \pi L^{^2}\Bigg) \Bigg].
\end{align}}\vspace{-0.15in}

\hspace{-0.15in}Since the PDF expression of $r_{_{\text{UL}}}$ is the same as that of $r_{_{\text{DL}}}$, the second part of the integrations in (\ref{A1}) is expressed as in (\ref{A3}). Based on that, the final expression of $\bar{d}_{_{\text{th}}}(L)$ is given by (\ref{dth}), which completes the proof.

\begin{figure}[t!]
  \centering
        \includegraphics[width=7cm,height=5cm]{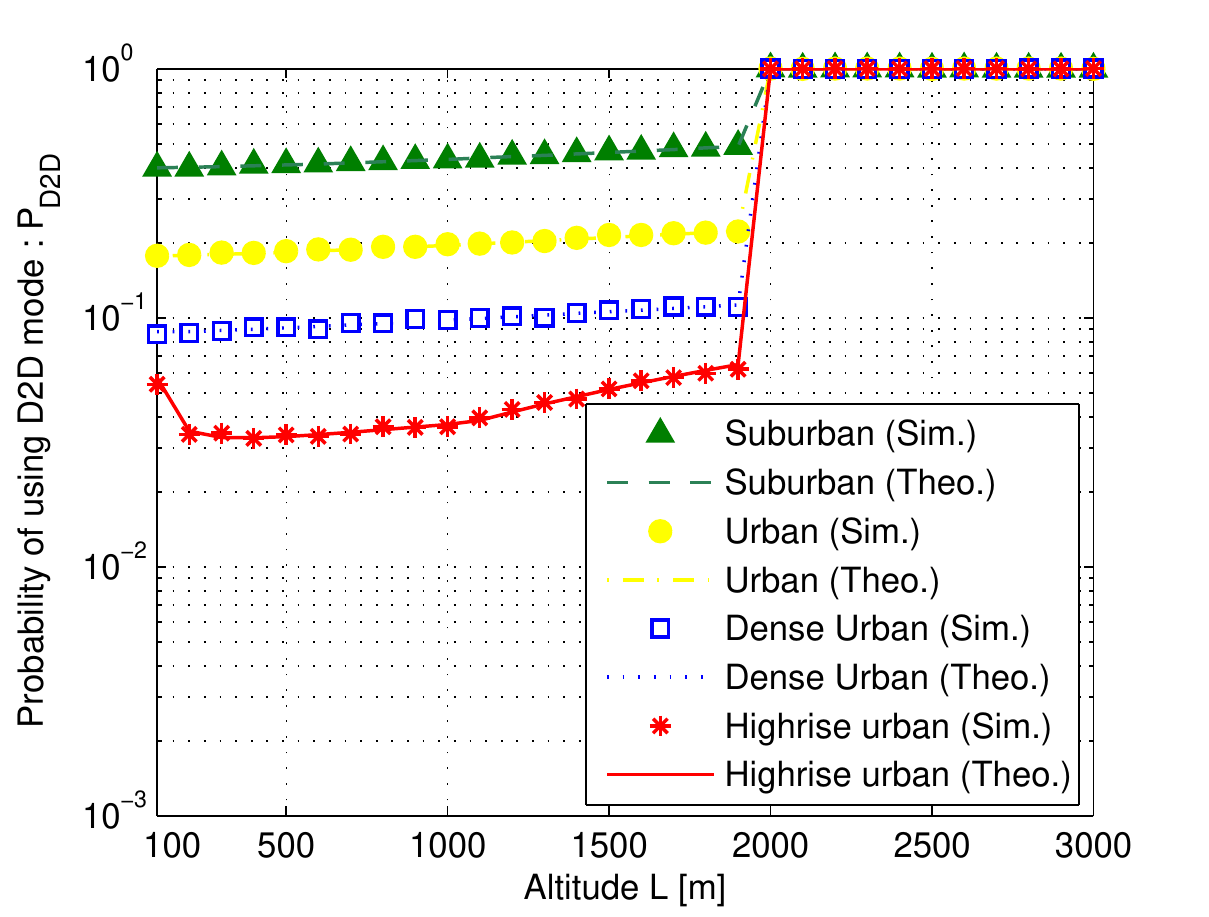}
				\vspace{-0.1in}
				    \caption{\footnotesize $P_{_{{D2D}}}$ vs. $L$, in different environments, with $\{\alpha, a,\ b,\ \eta_{_{\text{LOS}}},\ \eta_{_{\text{NLOS}}}\}$ are equal to $\{3.5,\ 27.23,\ 0.08,\ 2.3,\ 34\}$, $\{3.1,\ 12.08,\ 0.11,\ 1.6,\ 23\}$, $\{2.9,\ 4.88,\ 0.43,\ 1,\ 20\}$, and $\{2.7,\ 4.88,\ 0.43,\ 0.1,\ 21\}$ for high-rise urban, dense urban, urban, and suburban environments, respectively.}
						\vspace{-0.4in}
\end{figure} 

	\vspace{-0.1in}
\bibliographystyle{IEEEbib}
\bibliography{refs}

\end{document}